\documentclass[useAMS,usenatbib]{mn2e}

\usepackage{natbib}
\usepackage{graphicx}
\usepackage{amssymb}
\usepackage{color}
\usepackage{caption}

\title[AGN outflow energetics]{The energetics of AGN radiation pressure-driven outflows}

\author[ ]
{W. Ishibashi$^{1}$\thanks{E-mail:
wako.ishibashi@physik.uzh.ch}, A. C. Fabian$^{2}$ and R. Maiolino$^{3,}$$^{4}$
\footnotemark[0]\\
$^{1}$Physik-Institut, Universitat Zurich, Winterthurerstrasse 190, 8057 Zurich, Switzerland 
\footnotemark[0]\\
$^{2}$Institute of Astronomy, Madingley Road, Cambridge CB3 0HA \\
$^{3}$Cavendish Laboratory, University of Cambridge, 19 J. J. Thomson Ave., Cambridge CB3 0HE \\
$^{4}$Kavli Institute of Cosmology Cambridge, Madingley Road, Cambridge CB3 0HA
}

\begin{document}

\pdfminorversion=4


\date{Accepted 2018 January 24. Received 2017 December 31; in original form 2017 November 7}

\pagerange{\pageref{firstpage}--\pageref{lastpage}} \pubyear{2012}

\maketitle

\label{firstpage}

\begin{abstract} 
The increasing observational evidence of galactic outflows is considered as a sign of active galactic nucleus (AGN) feedback in action. However, the physical mechanism responsible for driving the observed outflows remains unclear, and whether it is due to momentum, energy, or radiation is still a matter of debate. The observed outflow energetics, in particular the large measured values of the momentum ratio ($\dot{p}/(L/c) \sim 10$) and energy ratio ($\dot{E}_k/L \sim 0.05$), seems to favour the energy-driving mechanism; and most observational works have focused their comparison with wind energy-driven models. Here we show that AGN radiation pressure on dust can adequately reproduce the observed outflow energetics (mass outflow rate, momentum flux, and kinetic power), as well as the scalings with luminosity, provided that the effects of radiation trapping are properly taken into account. In particular, we predict a sub-linear scaling for the mass outflow rate ($\dot{M} \propto L^{1/2}$) and a super-linear scaling for the kinetic power ($\dot{E}_k \propto L^{3/2}$), in agreement with the observational scaling relations reported in the most recent compilation of AGN outflow data.
We conclude that AGN radiative feedback can account for the global outflow energetics, at least equally well as the wind energy-driving mechanism, and therefore both physical models should be considered in the interpretation of future AGN outflow observations. 
\end{abstract}


\section{Introduction}

Active galactic nucleus (AGN) feedback is widely invoked in galaxy evolutionary scenarios, e.g. to reproduce the observed black hole-host galaxy correlations, but direct observational evidence has not been always clear-cut \citep[][and references therein]{Fabian_2012}. 
In recent years, a growing body of observational work has revealed the existence of powerful outflows on galactic scales, which are thought to provide the physical link connecting the small scales of the central black hole to the large scales of the host galaxy  
\citep{Sturm_et_2011, Maiolino_et_2012, Veilleux_et_2013, Spoon_et_2013, Cicone_et_2014, Carniani_et_2015, Feruglio_et_2015, Tombesi_et_2015, Gonzalez-Alfonso_et_2017, Fiore_et_2017}. 
These galactic outflows, often observed to extend on $\sim$kpc-scales, are typically characterised by high velocity ($v \sim 1000$km/s), high momentum flux ($\dot{p} \gtrsim 10 L/c$) and large kinetic power ($\dot{E}_k \sim 0.05 L$). The associated mass outflow rates can be quite high ($\dot{M} \sim 10^3 \mathrm{M_{\odot}/yr}$), implying short depletion timescales \citep{Sturm_et_2011, Cicone_et_2014}. The occurrence of such powerful outflows on galactic scales has often been interpreted as an observational proof of AGN feedback in action.  

However, the physics of the driving mechanism(s) remains unclear, and whether the observed outflows are powered by momentum, energy, or radiation is still a source of much debate \citep[e.g.][and references therein]{King_Pounds_2015}. 
One way of driving large-scale outflows is via quasi-relativistic winds launched from the vicinity of the central black hole, which generate shockwaves propagating into the host galaxy \citep{King_et_2011, Zubovas_King_2012, Faucher-Giguere_Quataert_2012}. In this scenario, two distinct regimes can be recognized, depending on whether the shocked wind can cool efficiently or not: `momentum-driving' at small radii and `energy-driving' at large radii. In the latter energy-driven regime, the large-scale AGN outflows are predicted to have momentum rates of $\dot{p} \sim 20 L/c$ and kinetic energy rates of $\dot{E}_k \sim 0.05 L$ \citep{Zubovas_King_2012}. A different mechanism for driving large-scale feedback is via radiation pressure on dust \citep{Fabian_1999, Murray_et_2005, Thompson_et_2015}. 
In this case, as the dust absorption cross section is much larger than the Thomson cross section ($\sigma_d/\sigma_T \sim 10^3$), the resulting coupling between AGN radiation field and the surrounding dusty gas can be greatly enhanced.   

At first sight, the observed outflow energetics, and in particular the large measured values of the momentum ratio ($\dot{p}/(L/c) \gtrsim 10$) and energy ratio ($\dot{E}_k/L \sim 0.05$), seem to favour the energy-driving mechanism, and apparently rule out direct radiation pressure-driving. We have previously argued that AGN radiation pressure on dust can potentially drive high-velocity outflows on $\sim$kpc scales, similar to the observed ones, provided that the effects of radiation trapping are taken into account \citep{Ishibashi_Fabian_2015}. Here we wish to compute the full energetics of AGN radiation pressure-driven outflows, by analysing the dependence on the underlying physical parameters, and compare our model results with the most up-to-date observational data reported in recent studies \citep[e.g.][]{Fiore_et_2017}. 

The paper is structured as follows. We first recall the basics of AGN radiative feedback and the significance of the effective Eddington limit (Section \ref{Sec_radiative_feedback}). We next compute the resulting outflow energetics: mass outflow rate, momentum flux, and kinetic power; alongside the derived quantities, momentum ratio and energy ratio; and analyse their dependence on the underlying physical parameters (Section \ref{Sec_outflow_energetics}). 
In Section \ref{Sec_comparison}, we compare our model predictions with observations available in the literature, and in particular the newly reported observational scaling relations. Finally, we consider the relation to other physical models (e.g. the wind energy-driving mechanism), and discuss the physical implications of AGN radiative feedback in the broader context of co-evolutionary scenarios (Section \ref{Sec_discussion}).


\section{AGN radiative feedback: radiation pressure on dust}
\label{Sec_radiative_feedback}

We consider AGN feedback driven by radiation pressure on dust, which sweeps up the surrounding material into an outflowing shell. 
We recall that the general form of the equation of motion is given by: 
\begin{equation}
\frac{d}{dt} [M_{sh}(r) v] = \frac{L}{c} (1 + \tau_{IR} - e^{-\tau_{UV}} ) - \frac{G M(r) M_{sh}(r)}{r^2}
\end{equation}
where $L$ is the central luminosity, $M(r)$ is the total mass distribution, and $M_{sh}(r)$ is the shell mass \citep{Thompson_et_2015, Ishibashi_Fabian_2015}. 
Here we consider the simple case of an isothermal potential ($M(r) = \frac{2 \sigma^2}{G} r$, where $\sigma$ is the velocity dispersion) and fixed-mass shell ($M_{sh}(r) = M_{sh}$), for which analytical limits can be derived, allowing us to gain some physical insight into the problem. 
The infrared (IR) and ultraviolet (UV) optical depths are given by:
\begin{equation}
\tau_{IR}(r) = \frac{\kappa_{IR} M_{sh}}{4 \pi r^2} 
\end{equation}
\begin{equation}
\tau_{UV}(r) = \frac{\kappa_{UV} M_{sh}}{4 \pi r^2} 
\end{equation}
where $\kappa_{IR}$=$5 \, \mathrm{cm^2 g^{-1} f_{dg, MW}}$ and $\kappa_{UV}$=$10^3 \, \mathrm{cm^2 g^{-1} f_{dg, MW}}$ are the IR and UV opacities, with the dust-to-gas ratio normalised to the Milky Way value (i.e. $\kappa_{IR}$=$5 \, \mathrm{cm^2 g^{-1}}$ and $\kappa_{UV}$=$10^3 \, \mathrm{cm^2 g^{-1}}$ for the Milky Way dust-to-gas ratio). 
Three distinct physical regimes can be identified according to the optical depth of the medium: optically thick to both IR and UV, optically thick to UV but optically thin to IR (single scattering limit), and optically thin to UV. 
The optical depth falls off with increasing radius as $\tau \propto 1/r^2$, and the corresponding IR and UV transparency radii are respectively given by: $R_{IR} = \sqrt{\frac{\kappa_{IR} M_{sh}}{4 \pi}}$ and $R_{UV} = \sqrt{\frac{\kappa_{UV} M_{sh}}{4 \pi}}$. 

A critical luminosity is obtained by equating the outward force due to radiation pressure to the inward force due to gravity, which can be considered as a generalised form of the Eddington luminosity ($L_E'$).  
The corresponding Eddington ratio is defined as: 
\begin{equation}
\Gamma = \frac{L}{L_E'} = \frac{L r^2}{c G M(r) M_{sh}(r)} (1 + \tau_{IR} - e^{-\tau_{UV}}) 
\end{equation}
In the case of the isothermal potential and fixed-mass shell, we recall that the Eddington ratios in the three optical depth regimes are respectively given by \citep[cf.][]{Ishibashi_Fabian_2016b}: 
\begin{equation}
\Gamma_{IR} = \frac{\kappa_{IR} L}{8 \pi c \sigma^2 r} 
\label{Eq_Gamma_IR}
\end{equation}
\begin{equation}
\Gamma_{SS} = \frac{L r}{2 c \sigma^2 M_{sh}} 
\label{Eq_Gamma_SS}
\end{equation}
\begin{equation}
\Gamma_{UV} = \frac{\kappa_{UV} L}{8 \pi c \sigma^2 r} 
\label{Eq_Gamma_UV}
\end{equation} 
We observe that the luminosity appears in all three regimes, while the dust opacity (or equivalently, dust-to-gas ratio) appears in the IR-optically thick and UV-optically thin regimes, but not in the single scattering limit. We also note that $\Gamma_{IR}$ and $\Gamma_{UV}$ are independent of the shell mass configuration, which is only relevant in the single scattering regime. The dependence of the effective Eddington ratio on the different physical parameters can be summarised as follows:
\begin{equation}
\Gamma_{IR} \propto \kappa_{IR} L \propto f_{dg} L 
\label{Eq_Gamma_IR_bis}
\end{equation}
\begin{equation}
\Gamma_{SS} \propto L/M_{sh}
\label{Eq_Gamma_SS_bis}
\end{equation}
\begin{equation}
\Gamma_{UV} \propto \kappa_{UV} L \propto f_{dg} L 
\label{Eq_Gamma_UV_bis}
\end{equation}

Solving the equation of motion (with a number of approximations), we obtain the analytic expression for the radial velocity profile of the outflowing shell: 
\begin{equation}
v(r) = \sqrt{\frac{2 L r}{c M_{sh}}  + \frac{\kappa_{IR} L}{2 \pi c R_0}} \, , 
\label{Eq_v}
\end{equation}  
where $R_0$ is the initial radius.
As the shell is accelerated outwards, the shell velocity will exceed the local escape velocity, and the outflowing shell can in principle escape the galaxy (but the actual outcome will depend on the details of the sweeping-up of ambient material and the temporal evolution of the central luminosity, cf Discussion). 


\section{Outflow energetics}
\label{Sec_outflow_energetics}

The basic physical quantities used in characterising the observed outflows are: the mass outflow rate, the momentum flux, and the kinetic power. 
In the observational works, the outflow energetics can be estimated as:
\begin{equation}
\dot{M}_{out} = \frac{M_{out}v}{R}
\end{equation}
\begin{equation}
\dot{P}_{out} = \dot{M}_{out} v
\end{equation}
\begin{equation}
\dot{E}_{out} = \frac{1}{2} \dot{M}_{out} v^2
\end{equation}
where $M_{out}$ is the mass of the outflowing gas. 
These values are computed in the so-called thin-shell approximation \citep[][and references therein]{Gonzalez-Alfonso_et_2017}; while in other studies a factor of 3 higher values are obtained by assuming a spherical geometry \citep[e.g.][]{Maiolino_et_2012, Fiore_et_2017}. 
Two derived quantities, the momentum ratio and the energy ratio, are often used to compare the observational measurements with model predictions: $\dot{P}_{out}/(L/c)$ and $\dot{E}_{out}/L$. 

It should be noted that $\dot{M}_{out}$, $\dot{P}_{out}$, and $\dot{E}_{out}$ are convenient snapshot parametrizations of a time-dependent process. In our calculations, $M_{out} = M_{sh}$ which is constant over time. Where $v$ is approximately constant at large times Êand radii, the definition of $\dot{M}_{out}$ means that it drops with increasing radius, similarly with the other parameters. The velocity $v$ is an integral Êresult Êof the earlier flow and this is lost in our snapshot definitions.


\subsection{Mass outflow rate, momentum flux, and kinetic power}
\label{subsec_3energetics}

By analogy with the observational works, we compute the corresponding model quantities characterising the outflow energetics: mass outflow rate ($\dot{M}$), momentum flux ($\dot{p}$), and kinetic power ($\dot{E}_k$): 
\begin{equation}
\dot{M} = \frac{M_{sh}}{t_{flow}} = \frac{M_{sh} v}{r}
\label{Eq_Mdot}
\end{equation} 
\begin{equation}
\dot{p} = \dot{M} v  = \frac{M_{sh} v^2}{r}
\label{Eq_pdot}
\end{equation} 
\begin{equation}
\dot{E}_k = \frac{1}{2} \dot{M} v^2 = \frac{1}{2} \frac{M_{sh} v^3}{r}
\label{Eq_Edot}
\end{equation}
We recall that here we simply follow the evolution of a single outflowing shell, and estimate the outflow energetics (i.e. the three quantities $\dot{M}, \dot{p}, \dot{E}_k$) in the thin-shell approximation, as adopted in the observational studies \citep[e.g.][]{Gonzalez-Alfonso_et_2017}.

\begin{figure}
\begin{center}
\includegraphics[angle=0,width=0.4\textwidth]{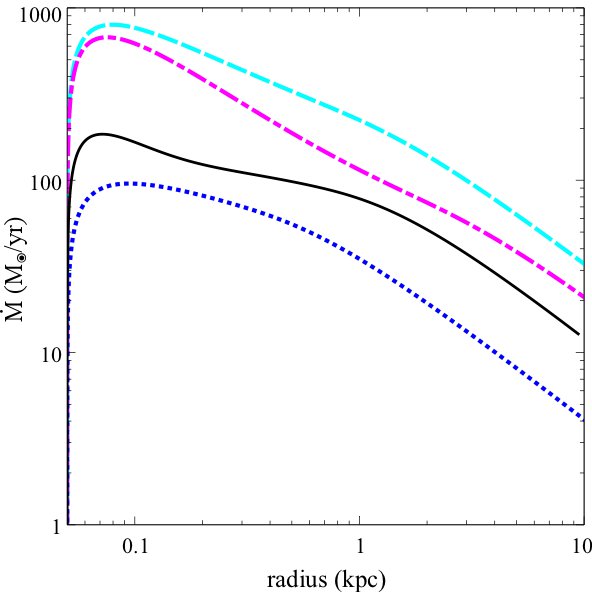} 
\caption{\small
Mass outflow rate vs. radius. $L = 10^{46}$erg/s, $M_{sh} = 10^8 M_{\odot}$, $f_{dg} = 1/150$, $R_0 = 50$pc (black solid). Variations: $L = 5 \times 10^{46}$erg/s (cyan dashed), $f_{dg} = 1/30$ (magenta dash-dotted), $M_{sh} = 2 \times 10^7 M_{\odot}$ (blue dotted). 
}
\label{Fig_Mdot_r}
\end{center}
\end{figure} 

\begin{figure}
\begin{center}
\includegraphics[angle=0,width=0.4\textwidth]{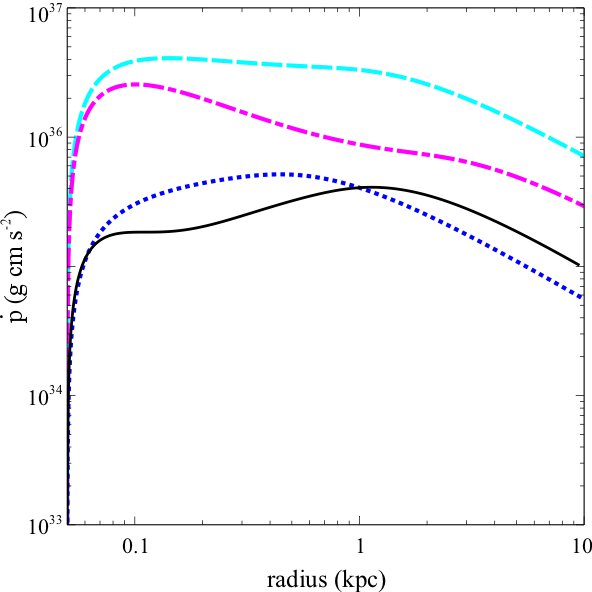} 
\caption{\small
Momentum flux vs. radius. $L = 10^{46}$erg/s, $M_{sh} = 10^8 M_{\odot}$, $f_{dg} = 1/150$, $R_0 = 50$pc (black solid). Variations: $L = 5 \times 10^{46}$erg/s (cyan dashed), $f_{dg} = 1/30$ (magenta dash-dotted), $M_{sh} = 2 \times 10^7 M_{\odot}$ (blue dotted). 
}
\label{Fig_pdot_r}
\end{center}
\end{figure} 

\begin{figure}
\begin{center}
\includegraphics[angle=0,width=0.4\textwidth]{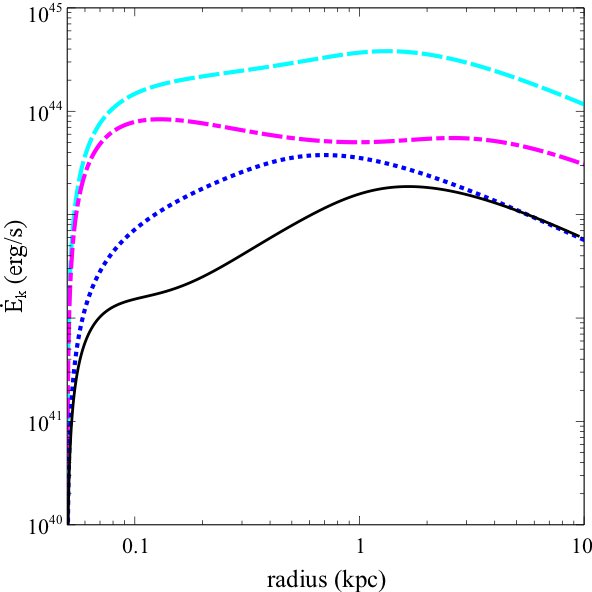} 
\caption{\small
Kinetic power vs. radius. $L = 10^{46}$erg/s, $M_{sh} = 10^8 M_{\odot}$, $f_{dg} = 1/150$, $R_0 = 50$pc (black solid). Variations: $L = 5 \times 10^{46}$erg/s (cyan dashed), $f_{dg} = 1/30$ (magenta dash-dotted), $M_{sh} = 2 \times 10^7 M_{\odot}$ (blue dotted). 
}
\label{Fig_Edot_r}
\end{center}
\end{figure}

In Figures  \ref{Fig_Mdot_r}, \ref{Fig_pdot_r}, and \ref{Fig_Edot_r}, we plot the mass outflow rate, momentum flux, and kinetic power, as a function of radius. Here the exact radial velocity profile, resulting from the full numerical integration, is used when computing the outflow parameters shown in the plots. The following values are taken as fiducial parameters of the model (black solid curve): $L = 10^{46}$erg/s, $M_{sh} = 10^8 M_{\odot}$, $f_{dg} = 1/150$, $R_0 = 50$pc, $\sigma = 200$km/s.  
We also consider variations by a factor of 5 in the physical parameters, by modifying one single parameter at a time while keeping the others fixed, in order to see which one has the major impact on the outflow energetics: enhanced luminosity (cyan dashed), reduced shell mass (blue dotted), and enhanced dust-to-gas ratio (magenta dash-dotted).  

In all three plots, we observe that the luminosity has the major effect in determining the outflow energetics, followed by the dust-to-gas ratio, and finally the shell mass. This trend may be qualitatively explained in terms of the dependence of the effective Eddington ratio on the underlying physical parameters (cf. Eqs. \ref{Eq_Gamma_IR_bis}-\ref{Eq_Gamma_UV_bis}): the luminosity appears in all three optical depth regimes, the dust-to-gas ratio in the IR-optically thick and UV-optically thin regimes, and $M_{sh}$ only in the single scattering regime. 
The exact location of the IR and UV transparency radii depend on the dust opacity and shell mass, being typically located around a hundred pc and a few kpc, respectively.  

Based on the definitions given in Eqs. (\ref{Eq_Mdot}-\ref{Eq_Edot}), we can derive the analytic limits for the mass outflow rate, momentum flux, and kinetic power: 
\begin{equation}
\dot{M}
=  \left( \frac{2 L M_{sh}}{c r}  + \frac{\kappa_{IR} L M_{sh}^2}{2 \pi c R_0 r^2} \right)^{1/2}
\end{equation} 
\begin{equation}
\dot{p}
= \frac{2 L}{c} + \frac{\kappa_{IR} L M_{sh}}{2 \pi c R_0 r}
\end{equation} 
\begin{equation}
\dot{E}_k 
= \frac{M_{sh}}{2r} \left( \frac{2 L r}{c M_{sh}} + \frac{\kappa_{IR} L}{2 \pi c R_0} \right)^{3/2} 
\end{equation} 

We note that the mass outflow rate scales with luminosity as $\dot{M} \propto L^{1/2}$, while the kinetic power scales with luminosity as $\dot{E}_k \propto L^{3/2}$. 


\subsection{Momentum ratio and energy ratio}
\label{subsec_xi_K}

We next consider the two derived quantities, the momentum ratio ($\zeta$) and the energy ratio ($\epsilon_k$), which can also be used to quantify the outflow energetics:
\begin{equation}
\zeta = \frac{\dot{p}}{L/c} 
\end{equation} 
\begin{equation}
\epsilon_k = \frac{\dot{E}_k}{L} 
\end{equation}

\begin{figure}
\begin{center}
\includegraphics[angle=0,width=0.4\textwidth]{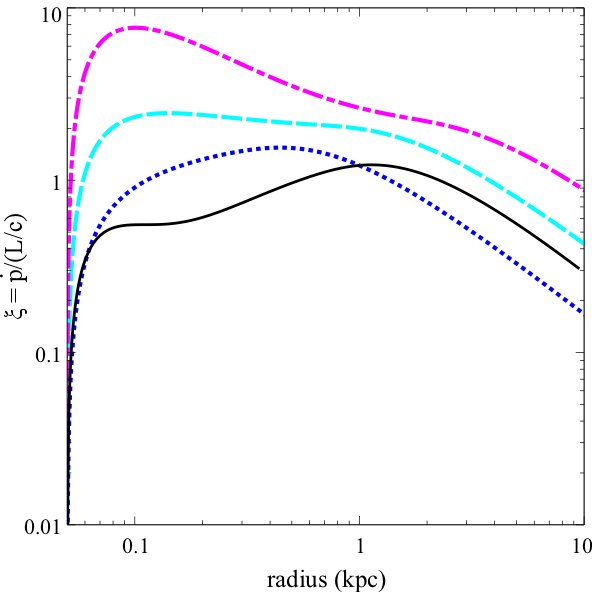} 
\caption{\small
Momentum ratio vs. radius. $L = 10^{46}$erg/s, $M_{sh} = 10^8 M_{\odot}$, $f_{dg} = 1/150$, $R_0 = 50$pc (black solid). Variations: $L = 5 \times 10^{46}$erg/s (cyan dashed), $f_{dg} = 1/30$ (magenta dash-dotted), $M_{sh} = 2 \times 10^7 M_{\odot}$ (blue dotted). 
}
\label{Fig_xi_r}
\end{center}
\end{figure}

\begin{figure}
\begin{center}
\includegraphics[angle=0,width=0.4\textwidth]{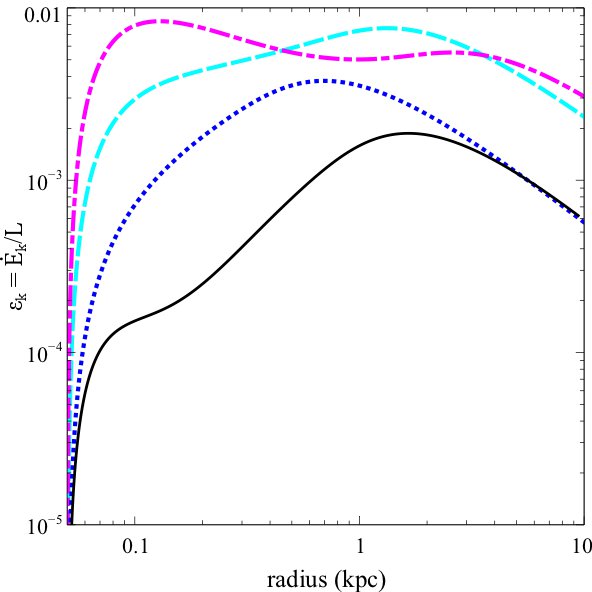} 
\caption{\small
Energy ratio vs. radius. $L = 10^{46}$erg/s, $M_{sh} = 10^8 M_{\odot}$, $f_{dg} = 1/150$, $R_0 = 50$pc (black solid). 
Variations: $L = 5 \times 10^{46}$erg/s (cyan dashed), $f_{dg} = 1/30$ (magenta dash-dotted), $M_{sh} = 2 \times 10^7 M_{\odot}$ (blue dotted). 
}
\label{Fig_K_r}
\end{center}
\end{figure}

Figures \ref{Fig_xi_r} and \ref{Fig_K_r} show the radial profiles of the momentum ratio and energy ratio corresponding to the shell models presented in Sect. \ref{subsec_3energetics}.  
As before, we can derive the analytic limits for the momentum ratio and the energy ratio: 
\begin{equation}
\zeta = 2 + \frac{\kappa_{IR} M_{sh}}{2 \pi R_0 r}
\label{Eq_xi}
\end{equation} 
\begin{equation}
\epsilon_k = \frac{M_{sh}}{2Lr} \left( \frac{2 L r}{c M_{sh}} + \frac{\kappa_{IR} L}{2 \pi c R_0} \right)^{3/2} 
\end{equation}
We see that the momentum ratio is independent of the luminosity, whereas the energy ratio scales with luminosity as $\epsilon_k \propto L^{1/2}$. The latter scaling implies that the energy ratio should be higher in more luminous sources. 
As previously mentioned, variations in the luminosity and dust-to-gas ratio can have a major effect on the outflow energetics, while the shell mass seems to be the less influential parameter. In fact, a larger shell mass implies a lower velocity but also a higher mass outflow rate, and the resulting momentum and energy ratios are broadly similar.

Compared to Figure \ref{Fig_xi_r}, we note that the analytic limit of the momentum ratio given in Eq. (\ref{Eq_xi}) tends to over-estimate the actual numerical values, since the analytic expression of the velocity provides an upper limit (e.g. the logarithmic term is neglected in Eq. \ref{Eq_v}). On the other hand, the second term dominates at small radii in Eq. (\ref{Eq_xi}), and for $r \sim R_{IR}$:
\begin{equation}
\zeta_{IR} \sim \frac{\kappa_{IR} M_{sh}}{2 \pi R_0 R_{IR}}
\approx \sqrt{\frac{\kappa_{IR} M_{sh}}{\pi R_0^2}} = 2 \sqrt{\tau_{IR,0}} \, , 
\end{equation}
which is equivalent to the relation $\zeta_{IR} = \frac{M_{sh} v_{IR}^2}{R_{IR} L/c}$, where $v_{IR} = \sqrt{\frac{\kappa_{IR} L}{2 \pi c R_0}}$ is the velocity near the IR transparency radius \citep[cf][]{Thompson_et_2015}.
Thus the momentum ratio is primarily determined by the initial IR optical depth, and large values can only be obtained if the optical depth to the reprocessed IR radiation is much larger than unity at the launch radius ($\tau_{IR,0} \gg 1$). 
Similarly, the energy ratio on small scales can be approximated as:
\begin{equation}
\epsilon_{k,IR} \approx \sqrt{\tau_{IR,0}} \frac{v_{IR}}{c} \, . 
\end{equation}
Therefore, both the momentum ratio and the energy ratio are mainly governed by the efficiency of radiation trapping, scaling as $\propto \sqrt{\tau_{IR,0}}$.


\section{Comparison with observations}
\label{Sec_comparison}

As mentioned in the Introduction, increasing observational evidence is emerging for galactic outflows, detected in ionised, neutral, and molecular gas phases \citep{Cicone_et_2014, Carniani_et_2015, Feruglio_et_2015, Tombesi_et_2015, Gonzalez-Alfonso_et_2017, Fiore_et_2017}. Molecular outflows are of particular interest, as they carry the bulk of the outflowing mass and comprise the medium from which stars ultimately form. Observations of molecular outflows indicate that the mass outflow rates are typically in the range $\dot{M} \sim (10-10^3) \mathrm{M_{\odot}/yr}$, the momentum rates in the $\dot{p} \sim (10^{35}-10^{37})$ gcm$s^{-2}$ range, and the kinetic luminosities in the $\dot{E}_k \sim (10^{42}-10^{45})$ erg/s range \citep{Cicone_et_2014, Carniani_et_2015, Fiore_et_2017}.
The model results shown in Figs. \ref{Fig_Mdot_r}-\ref{Fig_Edot_r} are broadly consistent with the observed numerical ranges, suggesting that AGN radiative feedback is potentially able to reproduce the global outflow energetics (a more detailed comparison is presented below). 
The most recent compilation of AGN outflows, obtained by collecting all available data from the literature, has been recently presented in \citet{Fiore_et_2017}. In the following, we focus our comparison with the energetics of molecular outflows reported in their sample, recalling that the quoted values should be divided by a factor of 3 to account for the difference in the assumed geometry. 

Observations indicate that molecular outflows typically have momentum ratios in the range $\zeta \sim (3-100)$, with half of the sources having momentum loads $> 10$ \citep{Fiore_et_2017}. Dividing by a factor of 3, the momentum ratio would be in the range $\zeta \sim (1-30)$, with typical values of a $\sim$few. From Figure \ref{Fig_xi_r}, we see that the predicted values of the momentum ratios are somewhat lower than the observed range, and in particular we cannot account for the highest $\zeta$ values. 
Concerning the energy ratio, molecular outflows are reported to have values in the range $\epsilon_k \sim (1-10) \%$, with an average ratio of $\sim 2.5\%$. Again dividing by a factor of 3, this implies that the energy ratio is typically in the range $\epsilon_k \sim (0.3-3) \%$, with an average value of $\sim 0.8\%$. Comparing with Figure $\ref{Fig_K_r}$, we note that the model energy ratios may account for the lower end of the observed range, but values exceeding $\epsilon_k > 0.01$ cannot be reproduced. 

\begin{figure}
\begin{center}
\includegraphics[angle=0,width=0.4\textwidth]{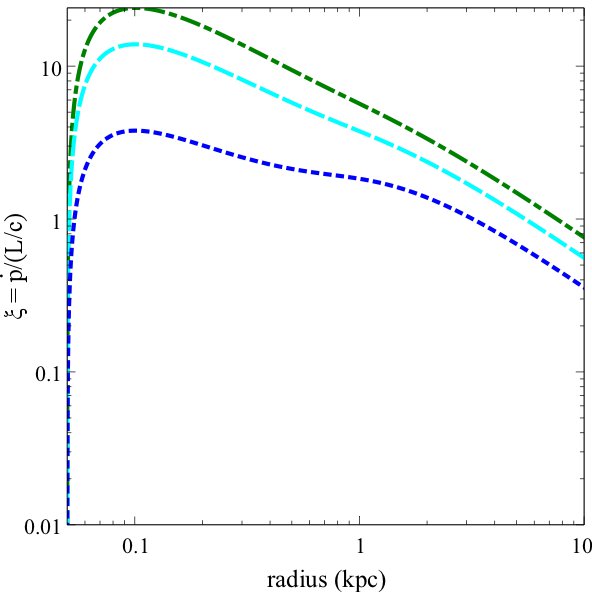} 
\caption{\small
Momentum ratio vs. radius for variations in the initial IR optical depth: $\tau_{IR,0} = 10$ (blue dotted), $\tau_{IR,0} = 30$ (cyan dashed), $\tau_{IR,0} = 50$ (green dash-dotted). 
}
\label{Fig_xi_r_var}
\end{center}
\end{figure} 
\begin{figure}
\begin{center}
\includegraphics[angle=0,width=0.4\textwidth]{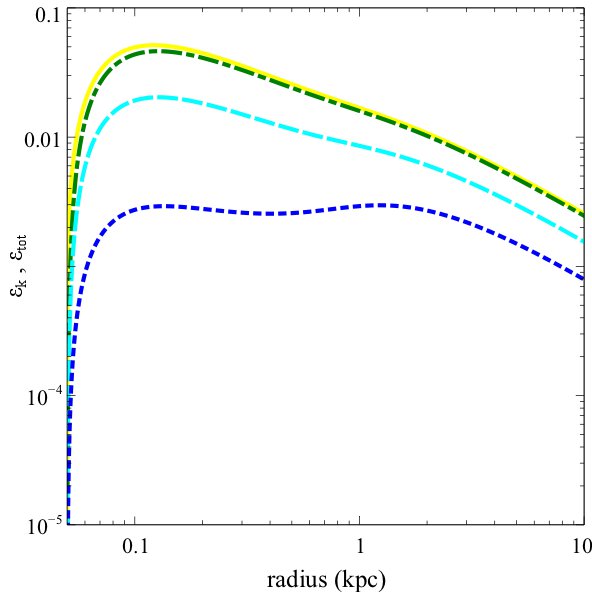} 
\caption{\small
Energy ratio vs. radius for variations in the initial IR optical depth: $\tau_{IR,0} = 10$ (blue dotted), $\tau_{IR,0} = 30$ (cyan dashed), $\tau_{IR,0} = 50$ (green dash-dotted). $\epsilon_{tot} = \epsilon_k + \epsilon_g$ (yellow solid), corresponding to the highest $\tau_{IR,0}$ model. 
}
\label{Fig_K_r_var}
\end{center}
\end{figure}

From the analysis in the previous section (Sect. \ref{subsec_xi_K}), it follows that the key parameter governing the outflow energetics is the initial IR optical depth. In order to evaluate the quantitative importance of this parameter, we plot the momentum ratio and energy ratio for enhanced IR optical depths (Figures \ref{Fig_xi_r_var} and \ref{Fig_K_r_var}). Large optical depths (due to high densities and large dust content) may be easily reached in the nuclear regions of obscured AGNs and ULIRG-like systems. 
As expected, we see that significantly higher values of the momentum ratio ($\zeta \gtrsim 10$) and energy ratio ($\epsilon_k > 0.01$) can now be obtained, which better reproduce the upper end of the observed range. 
Moderate values of $\zeta \sim 5$ and $\epsilon_k \lesssim 0.01$ are obtained on $\sim$kpc scales, consistent with the observational values, typically measured at radii $R \lesssim 1$kpc in molecular outflows \citep{Cicone_et_2014, Fiore_et_2017}. In our picture, the maximal values of the momentum and energy ratios are attained at small radii ($r \lesssim R_{IR}$), where the shell is optically thick to the reprocessed IR radiation. Thus efficient radiation trapping is required in order to account for the highest values of the momentum and energy ratios. In fact, the observed outflow energetics can potentially allow us to put some constraints on the physical conditions of the innermost regions of AGNs. 

\begin{figure}
\begin{center}
\includegraphics[angle=0,width=0.4\textwidth]{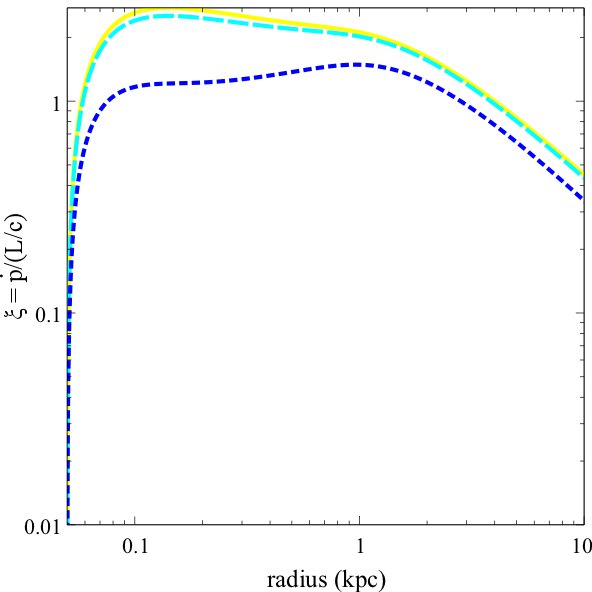} 
\caption{\small
Momentum ratio vs. radius for variations in luminosity and velocity dispersion: $L = 10^{46}$erg/s and $\sigma = 170$km/s (blue dotted), $L = 10^{47}$erg/s and $\sigma = 260$km/s (cyan dashed), $L = 10^{47}$erg/s and $\sigma = 170$km/s (yellow solid). 
}
\label{Fig_xi_r_varMsigma}
\end{center}
\end{figure}

\begin{figure}
\begin{center}
\includegraphics[angle=0,width=0.4\textwidth]{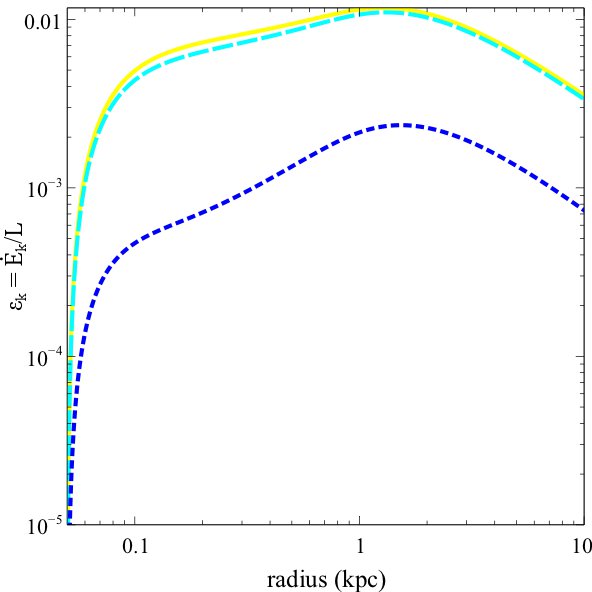} 
\caption{\small
Energy ratio vs. radius. Same parameters as in Fig. \ref{Fig_xi_r_varMsigma}. 
}
\label{Fig_K_r_varMsigma}
\end{center}
\end{figure}

For completeness, we also include the contribution of the work done against gravity: $W_g = \int \frac{G M(r) M_{sh}(r)}{r^2} dr = 2 \sigma^2 M_{sh} \ln \frac{r}{R_0}$, with the resulting gravitational ratio defined as $\epsilon_g = \dot{W}_g/L = 2 \sigma^2 M_{sh} v/ Lr$ (yellow solid curve in Figure \ref{Fig_K_r_var}). We see that the gravitational contribution seems to be unimportant in this particular case. 
Similarly, we can consider the effects of varying the velocity dispersion $\sigma$. Figures \ref{Fig_xi_r_varMsigma} and \ref{Fig_K_r_varMsigma} show the radial profiles of the momentum ratio and energy ratio, assuming a $M_{BH} - \sigma$ relation \citep[e.g.][]{McConnell_Ma_2013}, for two black holes of mass $M_{BH} \sim 10^8 M_{\odot}$ (blue dotted) and $M_{BH} \sim 10^9 M_{\odot}$ (cyan dashed), radiating at their respective Eddington luminosities. We recall that the effective acceleration is given by $a = \frac{L}{c M_{sh}} \left( 1 + \tau_{IR} - e^{-\tau_{UV}} \right) - \frac{2 \sigma^2}{r}$, which may be written as $a = a_{rad} + a_{grav}$. In general, the outflow propagation is facilitated in shallower potential wells. However, we note that for high enough luminosities, the acceleration is entirely dominated by the driving term ($a_{rad}$), and variations in the velocity dispersion (within a plausible $\sigma$ range) have not much influence on the outflow energetics (the cyan dashed and yellow solid curves almost overlap in Figs. \ref{Fig_xi_r_varMsigma} and \ref{Fig_K_r_varMsigma}). 

Observations also indicate that the mass outflow rate and kinetic power are well correlated with the AGN luminosity \citep{Fiore_et_2017}. The observational scalings for molecular outflows are given by: 
\begin{equation}
\dot{M} \propto L^{0.76\pm0.06}
\end{equation}
\begin{equation}
\dot{E}_k \propto L^{1.27\pm0.04}
\end{equation}
In our picture, we naturally expect a correlation between the outflow properties and the central luminosity (as the luminosity is the main parameter governing the effective Eddington ratio). More precisely, we derive that the mass outflow rate and kinetic power scale with luminosity as $\dot{M} \propto L^{1/2}$ and $\dot{E}_k \propto L^{3/2}$, respectively (Section \ref{subsec_3energetics}). 
We note that the theoretical scalings derived from the analytical limits are quite close to the observational scaling relations (also given the large uncertainties in the observational measurements). In particular, we predict a sub-linear scaling for the mass outflow rate and a super-linear scaling for the kinetic power, in agreement with the observational results. The latter scaling also implies that the energy ratio should scale with luminosity as $\epsilon_k \propto L^{1/2}$ (Sect. \ref{subsec_xi_K}). 
We further note that the energy ratio at small radii increases with increasing shell mass and decreasing initial radius (roughly scaling as $\propto M_{sh}^{1/2} R_0^{-3/2}$).



\section{Discussion}
\label{Sec_discussion}

\subsection{Model assumptions}

Here we assume spherical symmetry (with high gas covering fraction), which should be a valid approximation, especially in the heavily enshrouded nuclei of buried quasars and ULIRG-like systems. In realistic situations, the reprocessed radiation may tend to leak out through lower density channels, and the rate of momentum transfer may be reduced. Nonetheless, radiative transfer calculations, including multi-dimensional effects, indicate that values of several times $L/c$ can still be reached \citep{Roth_et_2012}. Even if the radiation-matter coupling is somewhat  reduced, compared to the case of a smooth spherical gas distribution, AGN radiative feedback due to the partial trapping of IR photons must still play a crucial role in initiating the outflow at early times. The actual efficiency of radiation trapping has been probed via different numerical simulations (see Section \ref{Subsec_radiation_trapping}). 

Substantial momentum and energy boosts can be obtained, provided that the optical depth to the reprocessed IR radiation is much larger than unity at the launch radius ($\tau_{IR,0} \gg 1$). Observations of ULIRGs indicate that huge amounts of gas, with very high column densities, are concentrated in the inner $\lesssim 100$pc region \citep[][and references therein]{Aalto_et_2015}. Such compact, buried nuclei can be optically thick to IR and even submm wavelengths.
In principle, a constraint on the initial radius can be derived from the observational measurements of the outflow energetics. We have previously tried such a test for the particular case of Mrk 231, obtaining a rather small initial radius of $R_0 \sim 10$pc \citep{Ishibashi_Fabian_2015}. However, major uncertainties are involved, especially in cases when the central AGN luminosity varies over time. A strict lower limit to the initial radius is only set by the dust sublimation radius, $R_{sub} = \sqrt{\frac{L}{16 \pi \sigma_{SB} T_{sub}^4}}$, which is of the order of $R_{sub} \sim 1$pc for typical parameters.

\subsection{Comparison with other forms of driving mechanisms}

The large observed values of the momentum ratio ($\zeta \sim 10$) and energy ratio ($\epsilon_k \sim 5\%$) have been interpreted as an indication that the outflows are in the energy-driven regime. Indeed, the outflow energetics apparently seems to favour `energy-driving' over `radiation pressure-driving', and most observational works have focused their comparisons with wind energy-driving models \citep[e.g.][]{Zubovas_King_2012}. Here we explicitly show that AGN radiation pressure on dust is capable of driving powerful outflows on galactic scales, and that high momentum and energy ratios can be reproduced, provided that the reprocessed radiation is efficiently trapped in the inner regions. Moreover, the observational scalings of the mass outflow rate and kinetic power can be naturally accounted for in our radiative feedback scenario. Hence, by properly taking into account the effects of radiation trapping, AGN radiative feedback is able to explain the observed outflow energetics, at least equally well as wind energy-driven models.  

In the case of the wind outflow model, the inner wind is assumed to be launched from the immediate vicinity of the central black hole, with a mass rate comparable to the Eddington rate (with $\dot{m} =  \dot{M}_w/\dot{M}_E \sim 1$) \citep{Zubovas_King_2012, King_Pounds_2015}. The resulting large-scale outflows are expected to have kinetic luminosities of $\dot{E}_k \sim \frac{\eta}{2 \dot{m}} L \sim 0.05 L$, where $\eta \sim 0.1$ is the standard accretion efficiency.
The quoted value of $\sim 5\%$ is often compared with the observational measurements of galactic outflows \citep[e.g.][]{Cicone_et_2014}. If taken at face value, this would imply a fixed coupling efficiency, with the energy ratio being basically set by the accretion efficiency (but see below for a potential dependence on $\dot{m}$). Since the accretion efficiency is determined by the black hole spin parameter, it then follows that the energy ratio should be a monotonic function of black hole spin (which may not have a straightforward physical interpretation). 
Actually, the global change in internal energy is given by the energy injection rate minus the rate of PdV work and the work against gravity, and the overall coupling efficiency may be further reduced in the case of leaky shells \citep[][and references therein]{King_Pounds_2015}.

The fact that the observed $\epsilon_k$ values are close to the predicted $\sim 5\%$ has often been taken as evidence for the energy-driving mechanism. But a closer inspection suggests that the observational values mostly tend to lie below the canonical $\sim 5\%$ line \citep{Cicone_et_2014, Carniani_et_2015, Fiore_et_2017}. Within the wind outflow scenario, it has been argued that lower values of the momentum and energy loading factors might be preserved, if the AGN luminosity evolution follows a power-law decay \citep{Zubovas_2018}. On the other hand, lower values of $\epsilon_k$ can be obtained by assuming $\dot{m} > 1$, for a given standard accretion efficiency. This would require some form of super-Eddington ejection. Although super-Eddington flows may  occur in stellar-mass black holes in binary systems (observed as ULXs), they may not hold for super-massive black holes, which tend to stay near-Eddington \citep{King_Muldrew_2016}. Moreover, it is also possible that most ULXs\footnote{We note that no ULX has yet been confirmed as a black hole, whereas several are found to be neutron stars \citep[][and references therein]{Walton_et_2018}.} are powered by accreting neutron stars rather than black holes. 

In contrast, in our AGN radiative feedback scenario, the energy ratio explicitly depends on the different physical parameters of the source, such as the luminosity and the optical depth of the medium, leading to a range of possible $\epsilon_k$ values. We also expect that the coupling efficiency should be higher in high-luminosity systems, consistent with the observed super-linear correlation.
Furthermore, we would naturally expect lower $\epsilon_k$ values for moderate radiation trapping (Fig. \ref{Fig_K_r}). 
 
Interestingly, recent observational studies of molecular outflows suggest that discriminating between energy-driving and momentum-driving is not always trivial. For instance, re-analysis of the nearby ULIRG (F11119+3257) with new ALMA data, indicates that the large-scale CO outflow is not inconsistent with momentum-driving, and thus AGN radiation pressure cannot be ruled out \citep{Veilleux_et_2017}. Another example is the recently discovered UFO/BAL quasar at $z \sim 3.9$ (APM08279+5255), which presents momentum boosts also consistent with momentum-driven flows \citep{Feruglio_et_2017}. 
Furthermore, ongoing analysis of molecular outflows, selected in an unbiased way from the ALMA archive data, suggest on average lower momentum and kinetic rates than in previous works (Flutsch et al. in preparation).
Therefore, the most recent observations tend to indicate lower values of the momentum and energy ratios, even more easily compatible with AGN radiative feedback, without the need to require extreme optical depths. On the other hand, a few sources present much higher values, with $\zeta \gg 10$ and $\epsilon_k \gg 0.05$, which cannot be easily accounted for, even in the energy-driven scenario.

In reality, the central luminosity varies with time, and if $L$ has dropped over time (between the initial launching of the shell and the current shell location), the inferred momentum and energy ratios may be over-estimated \citep[as previously discussed in][]{Ishibashi_Fabian_2015}. This could explain the very large values observed in some sources, which may be interpreted as signs of a past powerful AGN episode that has since faded. 
On the other hand, the shell may sweep up mass as it expands outwards, and the amount of swept-up material will determine the fate of the outflow: either the outflowing shell may completely escape the galaxy, or remain trapped in the outer halo and later fall back. 

\subsection{The importance of radiation trapping}
\label{Subsec_radiation_trapping}

An important aspect of the AGN radiative feedback model is the strength of the radiation-matter coupling, which depends on the degree of radiation trapping. The actual efficiency of radiation trapping has been investigated in numerical simulations, including radiation pressure on dust in extreme environments. Early results, based on the flux limited diffusion (FLD) approximation, suggested that the rate of momentum transfer cannot reach values much exceeding the single scattering limit, due to the development of radiative Rayleigh-Taylor (RT) instabilities \citep{Krumholz_Thompson_2013}. But this conclusion has been challenged by subsequent simulations, based on the more accurate variable Eddington tensor (VET) method, which indicate that there can be continuous acceleration of dusty gas, despite the development of RT instabilities in the flow \citep{Davis_et_2014}. This has been confirmed by updated studies comparing the different numerical schemes: indeed, the dusty gas can be accelerated to large scales, and the momentum transfer can be considerably amplified with respect to the single scattering value \citep{Tsang_Milosavljevic_2015, Zhang_Davis_2017}. Most recently, radiation-hydrodynamic simulations of radiation pressure-driven shells find that the boost factor is roughly equal to the IR optical depth as predicted (except at the highest optical depths), largely confirming our analytic picture (\citet{Costa_et_2018}, see also \citet{Costa_et_2017}). Therefore, we should be slowly moving towards a consensus recognising the importance of AGN radiation pressure on dust in driving large-scale outflows.

From the observational perspective, we recall that most outflow measurements are based on samples of local ULIRGs and QSOs \citep{Cicone_et_2014, Gonzalez-Alfonso_et_2017, Fiore_et_2017}. The nuclear regions of dense starbursts and obscured AGNs are characterised by high densities and large dust content \citep[e.g.][]{Aalto_et_2015}, implying high IR optical depths. Such ULIRG-like systems should form particularly favourable conditions for AGN radiative feedback. 
In fact, we have previously shown that even dense gas can potentially be disrupted in the IR-optically thick regime, and that an increase in the dust-to-gas ratio facilitates the shell ejection \citep{Ishibashi_Fabian_2016b}. 
Indeed, large amounts of dust imply heavy obscuration, but also powerful feedback. We have further discussed how our radiation pressure-driven shell models may be applied to the recently discovered populations of dusty quasars \citep{Ishibashi_et_2017}. These sources `in transition' are likely observed in the short-lived blow-out phase, transitioning from dust-obscured starbursts to unobscured luminous quasars \citep[e.g.][]{Banerji_et_2015}. 
In a broader context, we have proposed how such radiative feedback, which directly acts on the obscuring dusty gas, may provide a natural physical interpretation for the observed co-evolutionary sequence \citep{Ishibashi_Fabian_2016b}. 
Therefore AGN radiative feedback naturally fits in the global picture of `black hole-host galaxy co-evolution' scenarios.


\section{Conclusion}

Summarising, galactic outflows are now starting to be commonly observed, but the physical mechanism(s) responsible for their driving is still a matter of debate. Here we show that AGN radiation pressure on dust can account for the global outflow energetics (including large momentum and energy ratios) and the recently reported observational scaling relations. 
Furthermore, AGN radiation pressure on dust provides a physical mechanism for removing the obscuring dust cocoon, leading to a natural interpretation of the observed co-evolutionary path. 
Accordingly, AGN radiative feedback must be considered as a viable mechanism for driving galactic outflows, along with the wind energy-driving mechanism. We thus encourage future observations to compare the outflow measurements with both models to try to better understand the physical nature of galactic outflows.


\section*{Acknowledgements }

WI acknowledges support from the University of Zurich. 
ACF acknowledges support from ERC Advanced Grant 340442 and RM acknowledges Advanced Grant 695671.

  
\bibliographystyle{mn2e}
\bibliography{biblio.bib}


\label{lastpage}

\end{document}